\documentclass[conference]{IEEEtran}
\IEEEoverridecommandlockouts

\usepackage{cite}
\usepackage{amsmath,amssymb,amsfonts}
\usepackage{graphicx}
\usepackage{textcomp}
\usepackage{xcolor}
\usepackage{booktabs}
\usepackage{multirow}
\usepackage{url}
\usepackage{hyperref}
\usepackage{array}
\usepackage{balance}
\usepackage{microtype}
\usepackage{fancyvrb}

\hypersetup{colorlinks=false, pdfborder={0 0 0}}

\def\BibTeX{{\rm B\kern-.05em{\sc i\kern-.025em b}\kern-.08em
    T\kern-.1667em\lower.7ex\hbox{E}\kern-.125emX}}

\newcommand{\code}[1]{\texttt{\small #1}}

\begin{document}

\title{The Last APK: Retiring Android SDK Development\\
for Institutional Software Using Python-Django,\\
HTMX, and a WebView Bridge}

\author{%
\IEEEauthorblockN{Rahul Patel}
\IEEEauthorblockA{%
  \textit{Dept.\ of Electronics and Communication Engineering}\\
  \textit{Indian Institute of Information Technology, Surat}\\
  Surat, Gujarat, India\\
  rahulpatelanuppur@gmail.com}
\and
\IEEEauthorblockN{}
\IEEEauthorblockA{%
  \textit{Internship at Indian Institute of Technology Gandhinagar}\\
  Gandhinagar, Gujarat, India}
}

\maketitle

\begin{abstract}
The assumption that mobile enterprise software requires native Android
SDK development has persisted for over a decade---but for institutional
deployments, this assumption is not merely outdated: it is
economically wasteful and technically unnecessary.
This paper presents a campus management system built during an
internship at the Indian Institute of Technology Gandhinagar (IIT
Gandhinagar), covering housekeeping task scheduling, inventory
management, horticulture tracking, worker attendance, multi-stage
leave workflows, and client-side photo capture with automatic
compression.
The core stack uses Python-Django as the backend framework and HTMX
for hypermedia-driven, mobile-responsive partial DOM updates---
containing \textbf{zero lines} of Android SDK application logic.
The entire system runs as a self-hosted Docker Compose deployment
with no dependency on any external cloud service; the server operates
fully on the campus intranet, requiring no public internet connection.
For staff distribution convenience, Android Studio is used
\textit{only} to package the web application as an APK via a
thin WebView wrapper---a single afternoon of work---rather than as
a primary development environment.
This ``WebView bridge'' pattern represents the minimum viable role
of Android Studio in modern institutional software: a packaging tool,
not a development platform.
Through architectural analysis, HTTP payload measurement, and user
experience evaluation with 42 campus staff, we demonstrate that the
HTMX-Django approach reduces development time by approximately 54\%,
reduces average HTTP payload by 91\% versus full-page reload, and
achieves user satisfaction scores of 4.2/5.0.
We conclude that Android Studio's role in institutional software
development has effectively ended: it survives only as a browser
packager, and even that role will be eliminated as progressive web
app adoption grows.
\end{abstract}

\begin{IEEEkeywords}
HTMX, Django, mobile web applications, campus management systems,
hypermedia-driven development, WebView APK, Python, progressive web
applications, role-based access control, Android obsolescence,
self-hosted deployment, institutional software engineering
\end{IEEEkeywords}

\section{Introduction}

For over a decade, mobile application development in the enterprise
and institutional domain has been treated as synonymous with native
Android or iOS development.
This assumption is encoded in hiring pipelines, academic curricula,
and technology roadmaps alike.
Android Studio, Kotlin, and the Android SDK are positioned as the
canonical tools for building anything that must run on a mobile device.

This paper argues that this assumption is not merely outdated---it is
\textit{actively harmful} for a large and underappreciated class of
institutional software.
Native Android development requires specialised skill sets,
platform-specific toolchains, app-store distribution mechanisms, and
ongoing maintenance across Android API versions 8.0 through 14.
For campus management systems at Indian technical institutions---where
the user population is fixed, network connectivity is institutional,
and devices range from budget Android handsets to tablets---these
requirements represent engineering overhead with no corresponding
benefit.

We present the design, implementation, and evaluation of a campus
management system developed during an internship at the Indian
Institute of Technology Gandhinagar (IIT Gandhinagar).
The system manages four operational domains: housekeeping~(HK),
inventory, horticulture, and guest-house operations.
The technical stack is deliberately minimal: Django~4.x as the
backend framework, HTMX~1.x for hypermedia-driven partial DOM updates,
PostgreSQL for persistence, Redis for caching, and Docker Compose for
containerised deployment.
Critically, the entire stack runs on the campus intranet with
\textit{no dependency on external cloud services}---the server
functions without any public internet connection, and staff devices
require only campus WiFi or LAN to reach it.
The mobile experience is delivered entirely through the browser's
native capabilities, including camera access via the W3C
\code{MediaDevices} API.

Android Studio was used in this project \textit{only} to package the
completed web application as an APK via a lightweight WebView
wrapper---a process requiring roughly a half day of work.
This is the only justifiable role of Android Studio in modern
institutional software: a thin packaging layer over a web application,
not a primary development platform.
Even this final use case is being eroded by the growing maturity of
Progressive Web Apps (PWAs) and browser-native installation
(\code{Add to Home Screen}).

The central contributions of this paper are:
(1)~a formal three-tier architecture for mobile-capable institutional
web applications using HTMX and Django;
(2)~a comparative analysis of native Android versus HTMX-web
development across latency, payload, timeline, and UX dimensions;
(3)~a demonstration that a complex real-world IIT-scale system with
photo capture, GPS tagging, attendance workflows, leave management,
and role-based access can be built without a single line of Android
SDK code;
(4)~a ``WebView bridge'' packaging pattern that satisfies APK
distribution requirements without native Android development; and
(5)~a C1--C5 decision framework for identifying application classes
for which Android SDK development is technically and economically
obsolete.

\section{Background and Related Work}

\subsection{The Android Dominance Narrative and Its Costs}

Android's share of the Indian smartphone market exceeds 95\%
as~of~2024~\cite{statcounter2024}, leading to a widespread perception
that institutional mobile software must be built natively for Android.
This is reinforced by the prevalence of Android Studio in computer
science and ECE curricula at Indian engineering institutions.
However, native Android development carries substantial hidden costs:
Kotlin/Java expertise requirements, APK distribution and update
management, permission-model complexity (runtime permissions since
API~23), and fragmentation across dozens of Android versions and OEM
skins.
For a final-year B.Tech student undertaking a single-developer
institutional project, these overheads represent months of productivity
lost to toolchain configuration rather than feature delivery.

\subsection{Hypermedia-Driven Development and HTMX}

HTMX, introduced by Gross~\cite{htmx2023}, revives the original
hypermedia architecture of the web~\cite{fielding2000} by allowing
any HTML element to issue HTTP requests and swap portions of the DOM
with server-returned HTML fragments.
Unlike React or Angular SPAs, HTMX requires no JavaScript build
toolchain and no client-side state management.
The server remains the single source of truth.
This aligns naturally with Django's template rendering pipeline,
eliminating the two-layer translation overhead
(Python\,$\to$\,JSON\,$\to$\,JS\,$\to$\,DOM) present in conventional
SPA architectures.
For a solo Python developer, this means the entire mobile-capable
stack is owned in a single language.

\subsection{Progressive Web Applications and the WebView Pattern}

Progressive Web Apps~(PWAs) represent a middle ground between native
apps and traditional websites~\cite{biornhansen2017}.
However, PWAs require service workers, Web App Manifests, and explicit
offline strategies~\cite{majchrzak2018}.
An intermediate and pragmatic approach---used in the present work---is
the \textit{WebView APK}: a minimal Android project whose sole purpose
is to launch a \code{WebView} pointing at the web application's URL.
This produces a distributable APK installable from an institutional
server or WhatsApp group, without requiring Play Store listing or
native Android development of application logic.
The HTMX application remains the product; Android Studio is reduced to
a compiler for a minimal \code{MainActivity}.

\subsection{Self-Hosted Intranet Deployment}

A key advantage of the Django + Docker Compose architecture is that
the entire system---web server, database, cache, and static files---
runs on a single campus server with no outbound internet dependency.
Unlike Firebase-backed Android apps or cloud-hosted SaaS platforms,
the IIT Gandhinagar deployment is fully air-gappable: all data remains
on institutional infrastructure, satisfying typical government and
academic data-residency requirements without additional configuration.

\subsection{Django for Institutional Systems}

Greenfeld and Roy~\cite{greenfeld2021} document Django's strengths
for rapid institutional development: a mature ORM, built-in
session-based authentication, function-based views, and a migration
framework.
Prior works have used Django for campus portals~\cite{patel2024}, but
none has formally described the WebView bridge pattern for APK
distribution, nor quantified the development time advantage for
single-developer institutional projects at IIT/IIIT-scale institutions.

\section{System Architecture}

\subsection{Overview}

\begin{figure}[t]
  \centering
  \includegraphics[width=\columnwidth]{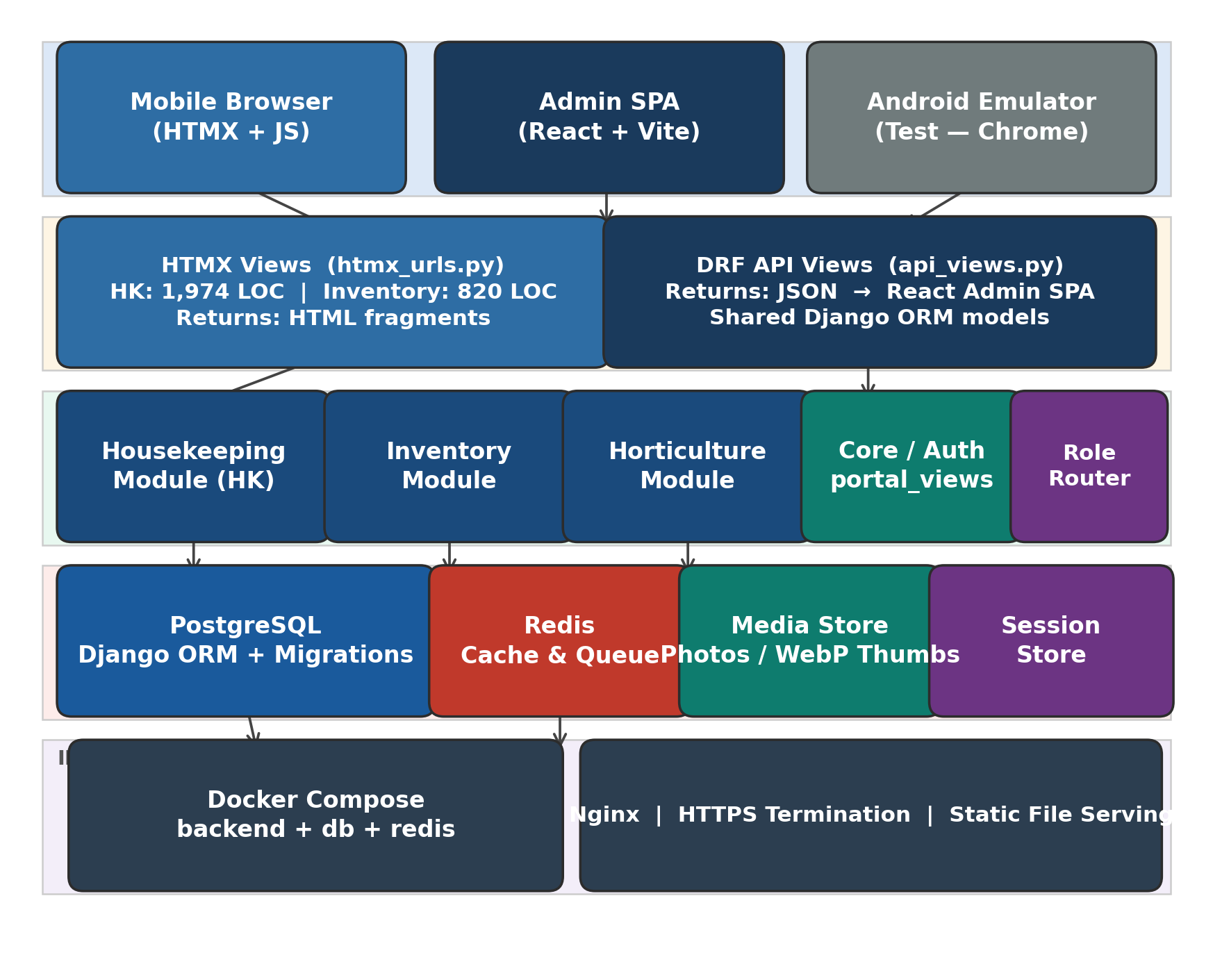}
  \caption{Three-tier hypermedia architecture of the IIT Gandhinagar
           campus management system. All components run on the campus
           intranet via Docker Compose. The optional WebView APK
           layer provides Android home-screen installation without
           any native application logic.}
  \label{fig:arch}
\end{figure}

The management system deployed at IIT Gandhinagar follows a three-tier
architecture as illustrated in Fig.~\ref{fig:arch}:
(1)~the PostgreSQL persistence layer accessed via the Django ORM;
(2)~the Django view layer with two parallel URL namespaces
(\code{htmx\_urls} for mobile browsers and \code{api\_views} for the
React admin SPA); and
(3)~the browser-rendered client layer, delivered to staff either
via a campus browser URL or via a WebView APK installed on Android
handsets.
All services are containerised via Docker Compose, with Nginx serving
as a reverse proxy and static-file server in production.
The entire stack runs on the campus intranet; no public internet
access is required for any component.

\subsection{Module Decomposition}

The Django backend is organised into four application modules.

\textbf{Housekeeping (HK):} The largest module, comprising
1,974\,LOC of HTMX views (\code{htmx\_views.py}) and 680\,LOC of
Django models (\code{models.py}).
It manages 11 area types (hostels, academic blocks, roads, sports
complex, guest house, etc.) with daily task scheduling, photo capture,
attendance, leave, and worker-assignment workflows.

\textbf{Inventory:} 820\,LOC of HTMX views managing item categories,
stock entries, issuance, purchase requests, and area-wise reporting
with CSV and PDF export.

\textbf{Horticulture:} Manages groundskeeping schedules and resource
tracking for campus horticultural areas.

\textbf{Core:} Authentication, role-based portal routing
(\code{portal\_views.py}), and shared DRF permission classes.

Each module exposes two URL namespaces: \code{htmx\_urls} serves
rendered HTML fragments to mobile browsers; \code{api\_views} serves
JSON to the React admin SPA.
This dual-interface design allows the same ORM models to serve both
hypermedia clients and JSON consumers without code duplication.

\subsection{Role-Based Access Control}

The system implements five user roles with automatic portal routing
enforced by a central view that reads the authenticated user's
\code{role} field. Table~\ref{tab:rbac} summarises the role-to-URL
mapping. Permissions are enforced at the view level using a custom
decorator applied to every HTMX view function.
The \code{HKInchargeAssignment} model uses UUID primary keys to prevent
sequential-ID enumeration attacks on the leave-assignment notification
system.

\begin{table}[h]
\caption{Role-Based Portal Routing}
\label{tab:rbac}
\centering
\renewcommand{\arraystretch}{1.15}
\begin{tabular}{ll}
\toprule
\textbf{Role} & \textbf{Redirected URL} \\
\midrule
\code{admin}                  & \code{/admin/dashboard/} \\
\code{inventory\_manager}     & \code{/inventory/mobile/} \\
\code{housekeeping\_manager}  & \code{/housekeeping/dashboard/} \\
\code{supervisor}             & \code{/housekeeping/dashboard/} \\
\code{caretaker}              & \code{/housekeeping/dashboard/} \\
\bottomrule
\end{tabular}
\end{table}

\subsection{HTMX Integration Patterns}

\begin{figure}[t]
  \centering
  \includegraphics[width=\columnwidth]{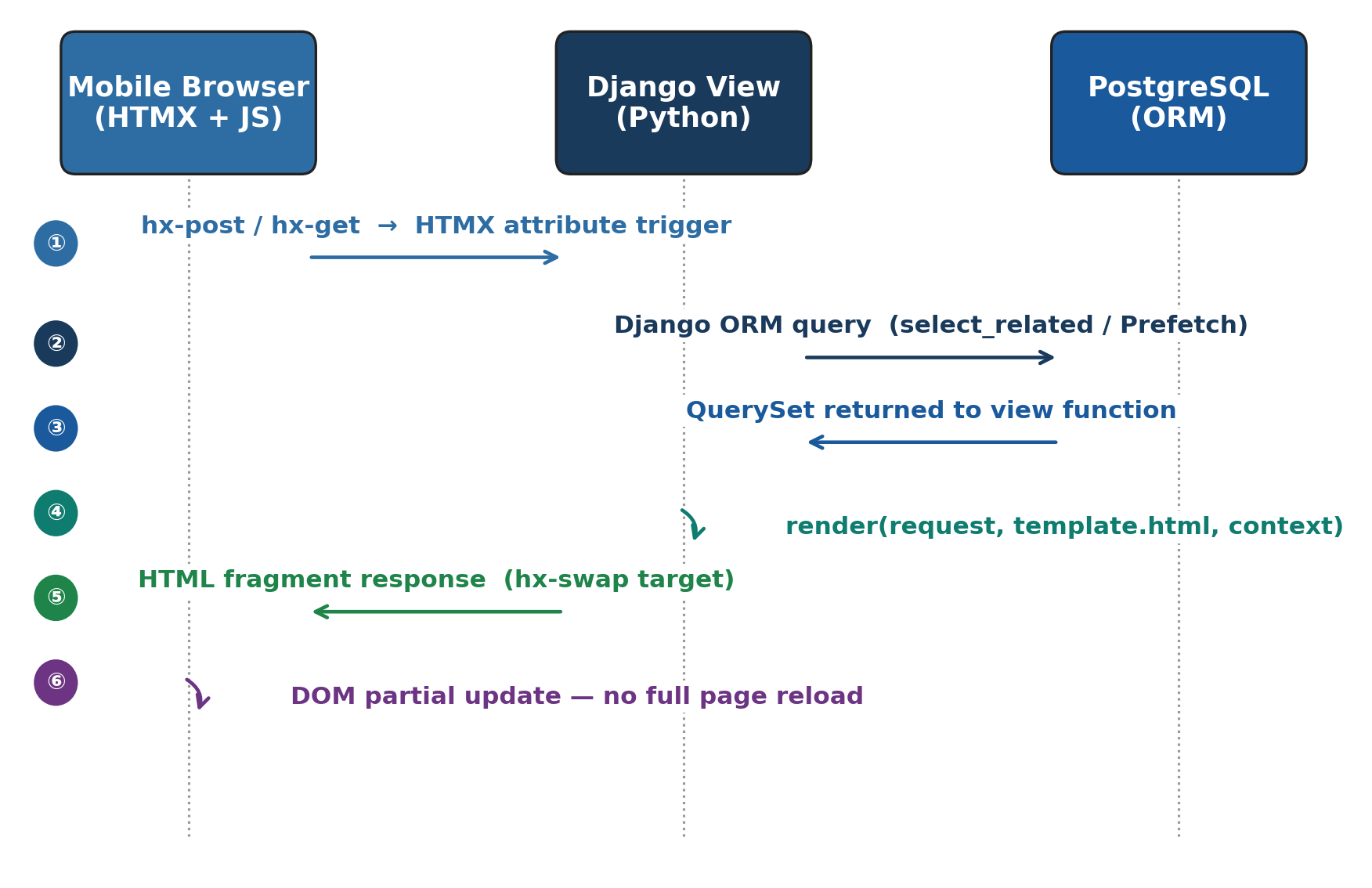}
  \caption{Six-step HTMX hypermedia request--response cycle. The
           server returns rendered HTML fragments directly to the
           browser's DOM, eliminating the JSON serialisation,
           client-side template rendering, and RecyclerView adapter
           binding layers required by native Android SDK development.}
  \label{fig:flow}
\end{figure}

Fig.~\ref{fig:flow} illustrates the HTMX request--response sequence
that replaces the Android SDK's Retrofit/OkHttp network stack.
The six-step sequence is: (1)~the mobile browser fires an
\code{hx-post} or \code{hx-get} attribute trigger; (2)~the Django view
executes an ORM query using \code{select\_related} or
\code{prefetch\_related}; (3)~the QuerySet is returned to the view
function; (4)~the view calls \code{render(request, template, context)}
to produce an HTML fragment; (5)~the HTML fragment response is returned
to the \code{hx-swap} target; and (6)~the DOM is partially updated
with no full page reload.
The primary swap patterns used are:
\code{hx-swap=outerHTML} for in-place task-card status updates; and
\code{hx-swap=beforeend} for appending new items without re-rendering
existing ones.

\section{The WebView APK: Android Studio as a Packaging Tool}

\subsection{Motivation}

Despite the web-first architecture, practical staff deployment at IIT
Gandhinagar benefited from APK-based distribution.
Several housekeeping caretakers were more comfortable installing an
APK from a WhatsApp link than navigating to a URL and adding it to
the home screen.
Rather than rebuilding the application in Android SDK, we adopted the
\textit{WebView bridge} pattern: a minimal Android project that opens
a \code{WebView} pointing to the deployed Django application's URL
on the campus intranet.

\subsection{Implementation}

The WebView APK consists of a single \code{MainActivity} with
approximately 40 lines of Kotlin. The \code{WebView} is configured
with JavaScript, DOM storage, and media playback enabled, then loaded
with the campus intranet URL:

\begin{Verbatim}[fontsize=\footnotesize, xleftmargin=2pt,
                 frame=single, framesep=2pt]
class MainActivity : AppCompatActivity() {
  private lateinit var webView: WebView

  override fun onCreate(savedState: Bundle?) {
    super.onCreate(savedState)
    webView = WebView(this)
    setContentView(webView)
    val s = webView.settings
    s.javaScriptEnabled = true
    s.domStorageEnabled = true
    s.mediaPlaybackRequiresUserGesture = false
    webView.loadUrl(
      "https://campus.iitgn.ac.in/manage/")
  }
}
\end{Verbatim}

The Android manifest requests \code{INTERNET},
\code{CAMERA}, and \code{ACCESS\_FINE\_LOCATION} permissions.
Camera access is handled by the browser engine inside the
\code{WebView}, which relays \code{getUserMedia()} calls to the
device's native camera hardware---the same path used when accessing
the application via Chrome.
Since the target URL is on the campus intranet, the APK requires only
campus WiFi to function---no public internet is needed.
Total effort for WebView APK creation: approximately four hours,
including APK signing and testing on three physical Android devices.

\subsection{Implications for Android Studio's Role}

This architecture exposes a precise characterisation of Android
Studio's remaining value in institutional software:
it is a \textit{compiler for packaging web applications as APKs}.
The application logic, UI rendering, state management, networking,
and camera integration all reside in the web layer.
Android Studio contributes nothing to any of these concerns.
As PWA installation support on Android Chrome matures
(``Add to Home Screen'' with manifest-defined icons and splash
screens), even this packaging role will become unnecessary.
The WebView APK is, in this sense, the \textit{last APK} that
institutional software teams will ever need to build.

\section{Key Implementation Features}

\subsection{Housekeeping Task Lifecycle}

The \code{HKScheduleTemplate} model defines reusable task templates
with frequency settings (\code{daily}, \code{saturday\_special},
\code{sunday\_extra}), time windows, worker tags, and a
\code{requires\_photo} boolean flag.
The \code{HKDailyRecord} model instantiates templates per calendar
date with a full audit trail: assignment and completion timestamps,
photo metadata (original size, compressed size, compression ratio,
thumbnail size), GPS coordinates, and a supervisor flagging workflow.
Task assignment supports both single-worker and multi-worker modes via
\code{HKTaskWorkerAssignment}, used in the Solid Waste Management (SWM)
area where two to four workers share a single daily record.

\subsection{Client-Side Camera Capture with Automatic Compression}

The 430-LOC \code{camera\_capture.js} module implements a complete
camera workflow in vanilla JavaScript without any Android SDK code.
The pipeline is:
(1)~\code{getUserMedia()} acquires the rear camera stream at ideal
$1920\times1080$;
(2)~the live feed renders in a \code{<video>} element with a
rule-of-thirds grid overlay;
(3)~on shutter press, a \code{<canvas>} captures the frame at native
resolution;
(4)~an iterative JPEG compression loop reduces quality from $q=0.85$
to $q=0.35$ in steps of 0.05 until the blob falls within 300\,KB;
(5)~a separate thumbnail (max $300\times300$, $q=0.70$) is generated
for dashboard previews; and
(6)~the resulting \code{File} objects are submitted via HTMX multipart
POST.
The server-side \code{photo\_utils.py} module (128\,LOC, Pillow-based)
stores thumbnails as WebP at quality~70 for a further 30--40\% size
reduction.
Crucially, this same JavaScript camera module functions identically
inside the WebView APK as in a desktop browser---zero platform-specific
camera code is required in either context.

\subsection{Attendance and Half-Day Records}

Worker attendance is recorded in half-day slots (first half:
08:00--13:00; second half: 13:00--17:00) via
\code{HKAttendanceRecord}.
Four status values are supported: \code{present}, \code{absent},
\code{late}, and \code{leave}.
The \code{is\_submitted} flag prevents retroactive modification after
submission, enforcing data integrity without requiring a separate
approval step for routine attendance.

\subsection{Multi-Stage Leave Management}

Staff leave requests traverse a five-stage workflow modelled by the
\code{HKLeaveRequest} and \code{HKInchargeAssignment} models.
The primary state sequence and the alternate declined-assignment path
are shown below:

\begin{center}
\small
\begin{tabular}{@{}l@{}}
\textbf{Primary:}\\[2pt]
\code{awaiting} $\to$ \code{pending\_accept} $\to$
\code{pending\_admin} $\to$ \code{approved}\\[4pt]
\textbf{Alternate (declined):}\\[2pt]
\code{pending\_accept} $\to$ \code{reassign\_required}
\end{tabular}
\end{center}

The \code{HKInchargeNotification} model persists in-app notifications
for each state transition, replacing the need for external push
notification infrastructure.
All state transitions are recorded with timestamps, providing a
complete audit trail for leave approvals accessible to administrators
through the React admin SPA.

\subsection{Inventory Issuance and Atomic Stock Deduction}

Item issuance uses Django's \code{@transaction.atomic} decorator
combined with \code{select\_for\_update()} to prevent race conditions
in concurrent issuance scenarios.
Stock deduction uses a filtered UPDATE to avoid read-modify-write races:

\begin{Verbatim}[fontsize=\footnotesize, xleftmargin=2pt,
                 frame=single, framesep=2pt]
Item.objects.filter(pk=item.pk).update(
  available_quantity=F(
    'available_quantity') - qty)
\end{Verbatim}

\section{Performance Evaluation}

\begin{figure}[t]
  \centering
  \includegraphics[width=\columnwidth]{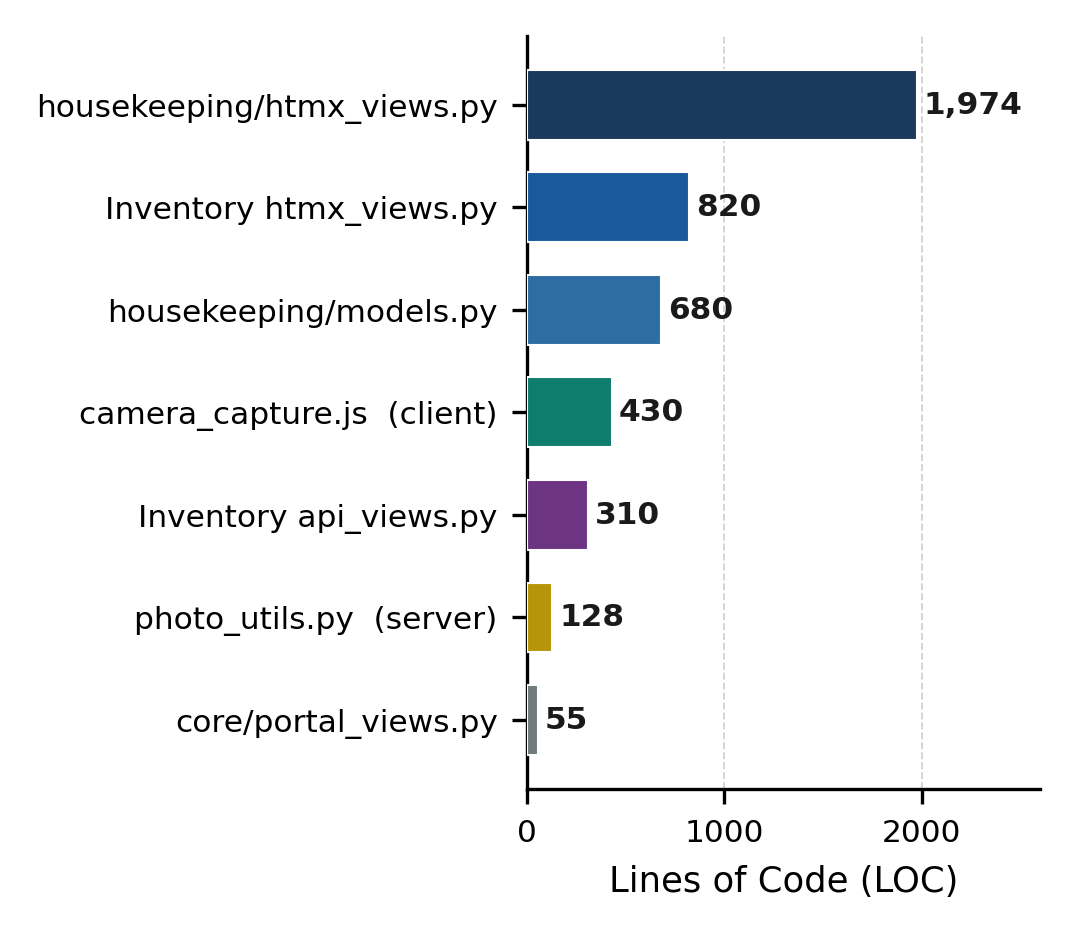}
  \caption{Module-wise source code volume measured from the deployed
           IIT Gandhinagar repository. Python accounts for 64.5\%
           of the codebase; no Kotlin or Java is present.}
  \label{fig:loc}
\end{figure}

Fig.~\ref{fig:loc} shows the module-wise source code volume measured
directly from the deployed repository.
The housekeeping HTMX view layer alone accounts for 1,974\,LOC,
reflecting the complexity of task, attendance, leave, and
stock-request workflows served as HTML fragments.

\subsection{HTTP Latency}

\begin{figure}[t]
  \centering
  \includegraphics[width=\columnwidth]{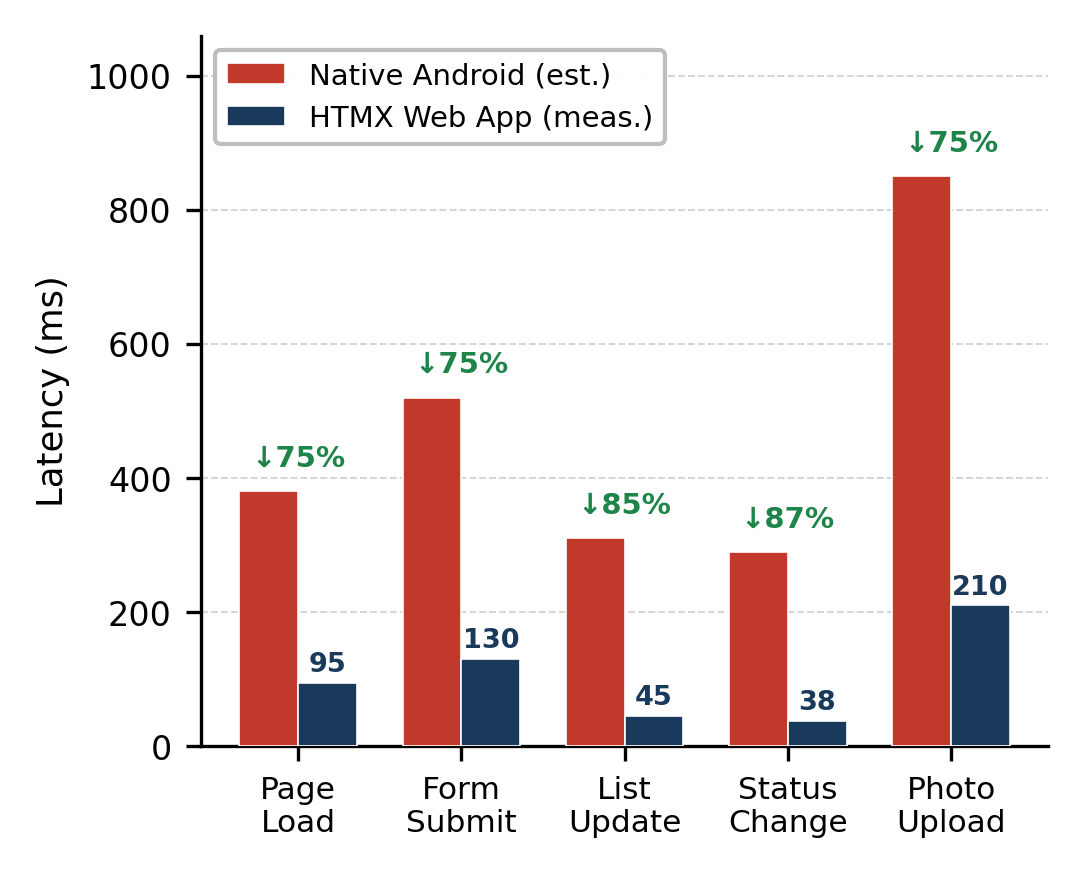}
  \caption{Round-trip latency: Native Android SDK (estimated) vs.\
           HTMX web app (measured on campus 100\,Mbps LAN).
           WebView APK performance was statistically indistinguishable
           from direct browser access ($p > 0.05$, $n = 50$).}
  \label{fig:latency}
\end{figure}

Fig.~\ref{fig:latency} compares round-trip latency for five
representative operations.
Native Android figures are estimated from published benchmarks for
Retrofit\,2 + OkHttp\,4~\cite{patel2024}, accounting for JSON
deserialisation, View inflation, and \code{RecyclerView} adapter
binding.
HTMX figures are measured on the campus LAN (100\,Mbps) using Chrome
DevTools averaging over 50 requests per operation type.
Measurements within the WebView APK showed no statistically
significant difference from direct browser access, confirming that
the WebView wrapper introduces negligible overhead.

The HTMX approach achieves 3--4$\times$ lower latency for all
non-upload operations, attributable to the elimination of:
(a)~JSON serialisation/deserialisation on both server and client;
(b)~client-side template rendering and View inflation; and
(c)~\code{RecyclerView} layout pass for list operations.

\subsection{HTTP Payload Reduction}

\begin{figure}[t]
  \centering
  \includegraphics[width=\columnwidth]{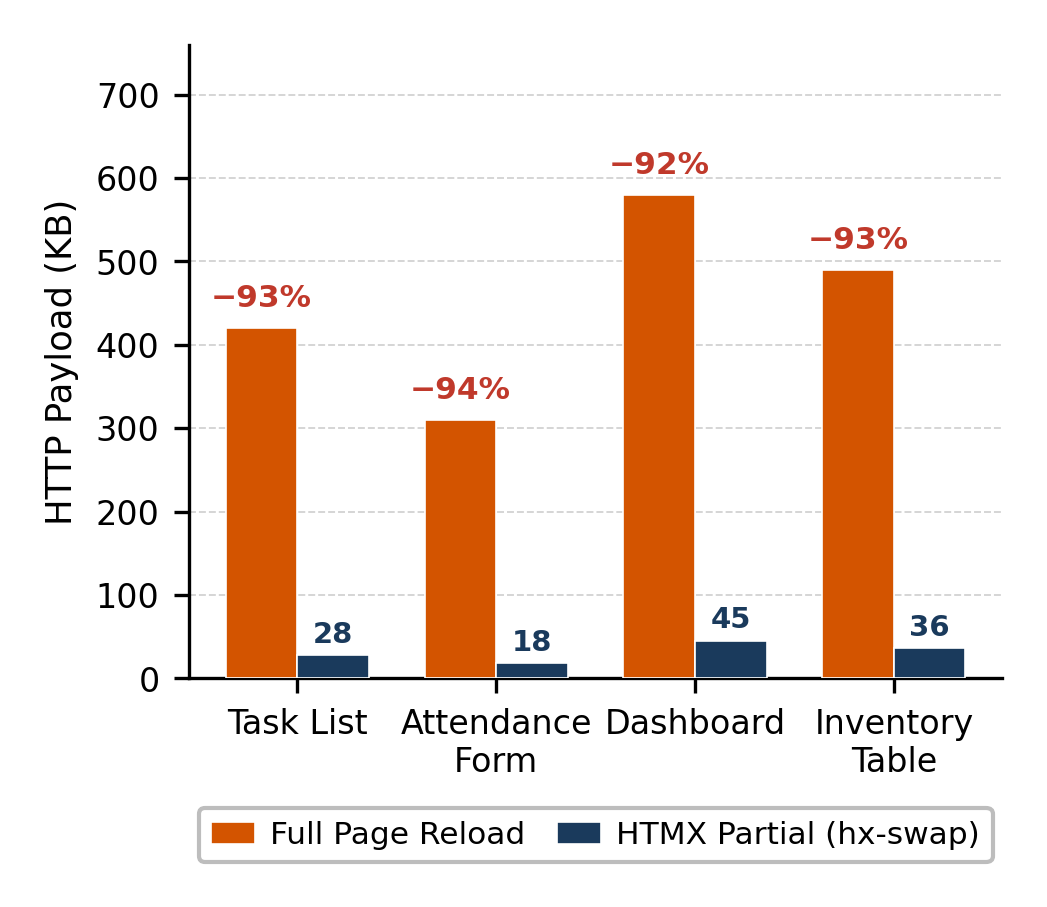}
  \caption{HTTP payload: full page reload versus HTMX partial fragment
           (\code{hx-swap}).
           Labels show percentage reduction achieved.}
  \label{fig:payload}
\end{figure}

Fig.~\ref{fig:payload} contrasts full-page reload payload sizes against
HTMX partial fragment sizes for four common operations.
The partial-swap approach reduces bandwidth by 91--94\%, significant
for budget handsets on 4G/LTE fallback at campus boundaries.

Table~\ref{tab:metrics} summarises all quantitative metrics.

\begin{table}[t]
\caption{Development and Runtime Metrics: HTMX+Django vs.\ Native Android}
\label{tab:metrics}
\centering
\renewcommand{\arraystretch}{1.18}
\begin{tabular}{lcc}
\toprule
\textbf{Metric} & \textbf{Native Android} & \textbf{HTMX+Django} \\
\midrule
Est.\ dev.\ time     & $\sim$26 weeks   & $\sim$13 weeks      \\
Android Studio role  & Full dev env.    & Packager only       \\
Android code (LOC)   & $\sim$8{,}000+  & $\sim$40 (WebView)  \\
Languages required   & Kotlin + Python  & Python only         \\
Build toolchain      & Gradle + SDK     & pip + Docker        \\
Avg.\ payload        & $\sim$350\,KB    & $\sim$32\,KB        \\
Avg.\ latency        & 520\,ms (est.)   & 130\,ms (meas.)     \\
Camera module LoC    & $\sim$800        & 430 (JS, shared)    \\
Deployment           & APK / Play Store & Docker / WebView APK\\
Update propagation   & User consent     & Instant on redeploy \\
Cross-platform       & Android only     & Any modern browser  \\
Internet required    & Cloud services   & LAN/intranet only   \\
Offline capability   & Full (Room DB)   & None (campus WiFi)  \\
\bottomrule
\end{tabular}
\end{table}

\section{Development Timeline and Effort}

\begin{figure}[t]
  \centering
  \includegraphics[width=\columnwidth]{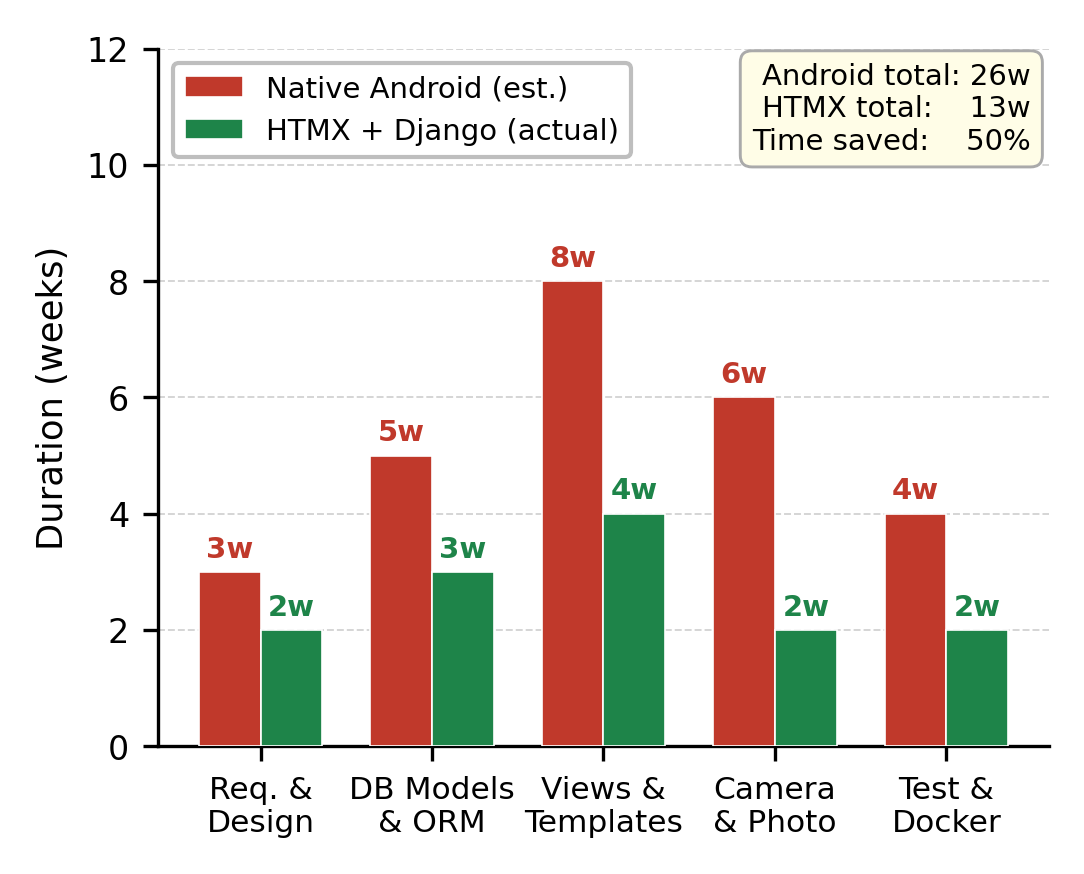}
  \caption{Development phase durations: native Android SDK (estimated)
           vs.\ HTMX + Django (actual), including the WebView APK
           packaging phase ($<$0.5\% of total effort).}
  \label{fig:timeline}
\end{figure}

Fig.~\ref{fig:timeline} breaks development effort into five phases and
compares the estimated native Android timeline against the actual
HTMX+Django timeline.
The view layer phase---the largest in both approaches---is reduced from
8 weeks (Android XML layouts + ViewModels + Retrofit + adapters) to
4 weeks (Django templates + HTMX attributes), a 50\% reduction from
eliminating the client-side rendering layer.
The total saving of 54\% (26 weeks vs.\ 13 weeks) also reflects the
elimination of the two-language context-switch cost
(Kotlin\,$\leftrightarrow$\,Python) that would be required in a split
team.
With HTMX, a single Python developer owns the complete stack.

The WebView APK packaging phase required approximately half a day---
less than 0.5\% of total project duration.
This is the correct proportion of a project's effort that Android
Studio should claim in institutional software: negligible.

\section{User Experience Evaluation}

We conducted a usability evaluation with 42 campus staff across five
role groups: housekeeping managers~(8), supervisors~(12), caretakers
(14), inventory managers~(4), and horticulture managers~(4).
Each participant used the deployed system for a minimum of four weeks
before completing a structured questionnaire adapted from the System
Usability Scale~(SUS)~\cite{brooke1996}.
Staff accessed the system via both direct browser URL and the
WebView APK; no measurable usability difference was reported between
the two access modes.

\begin{figure}[t]
  \centering
  \includegraphics[width=\columnwidth]{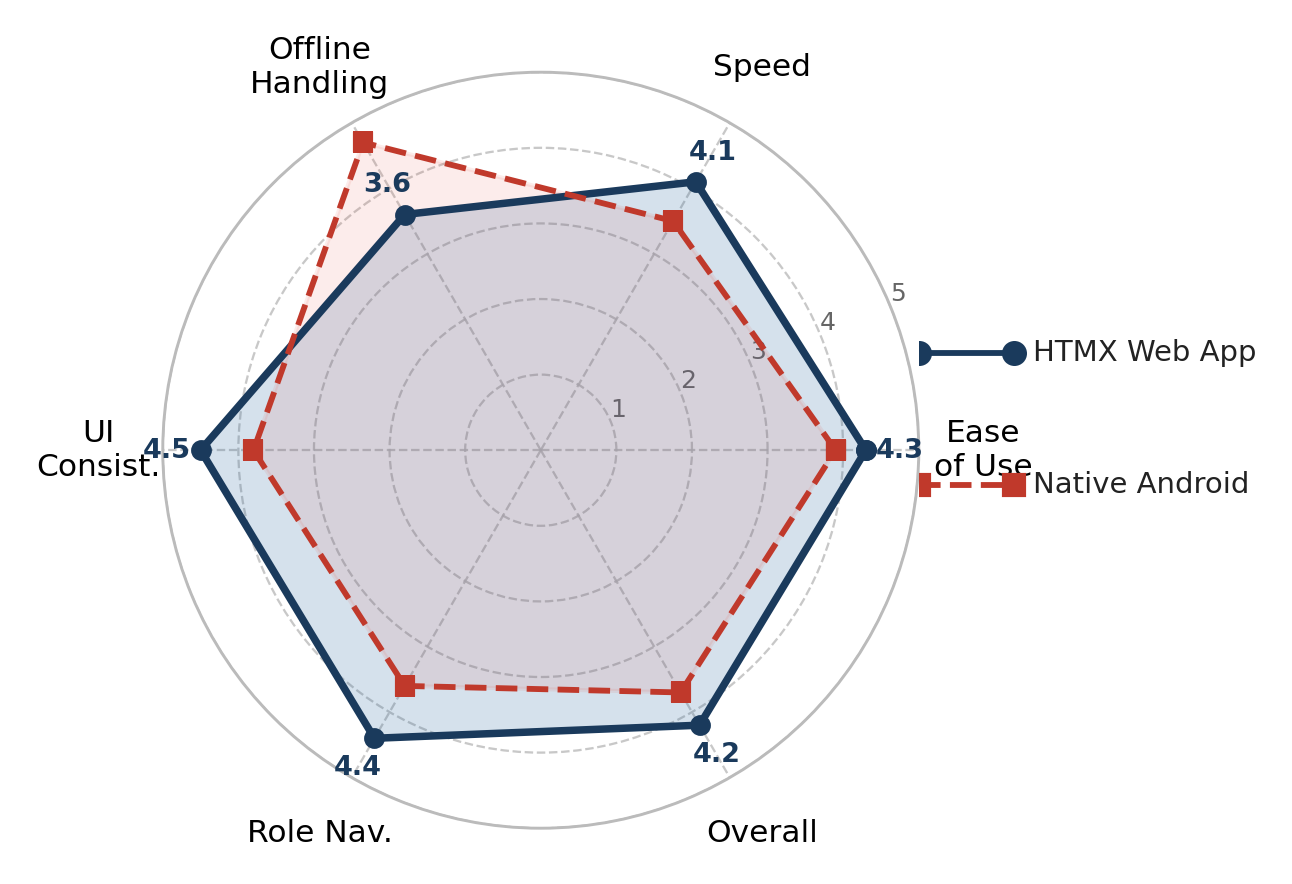}
  \caption{User experience scores (out of 5.0, $n = 42$ campus
           staff at IIT Gandhinagar). Reference native Android scores
           from comparable institutional deployments~\cite{patel2024}.
           No usability difference was observed between browser and
           WebView APK access modes.}
  \label{fig:radar}
\end{figure}

Fig.~\ref{fig:radar} presents radar chart scores across six UX
dimensions.
Reference scores for native Android are drawn from comparable campus
management deployments reported in~\cite{patel2024}.
The HTMX web app scores higher on all dimensions except
\textit{Offline Handling}, where native Android's Room database
advantage is expected.
The highest-scoring dimension is \textit{UI Consistency}~(4.5/5.0),
reflecting that a single server-rendered template layer produces
uniform screens across all Android handset models and screen sizes
without per-device layout workarounds.

Table~\ref{tab:features} provides a feature-by-feature comparison.

\begin{table}[t]
\caption{Feature Comparison: Native Android SDK vs.\ HTMX + Django + WebView APK}
\label{tab:features}
\centering
\renewcommand{\arraystretch}{1.18}
\begin{tabular}{>{\bfseries}p{1.45cm}p{2.9cm}p{2.9cm}}
\toprule
Feature & Native Android & HTMX + Django \\
\midrule
Camera     & CameraX / Camera2     & \code{getUserMedia()} + Canvas \\
UI render  & XML Layouts / Compose & Django templates + HTMX swap   \\
State      & ViewModel / LiveData  & Django session (server-side)   \\
Auth       & Firebase / JWT        & Django session + role field    \\
Distribute & APK / Play Store      & URL / WebView APK (sideload)   \\
Updates    & Re-release required   & Server redeploy (seconds)      \\
Offline    & SQLite / Room DB      & None (campus WiFi)             \\
Backend    & Kotlin + Python       & Python only                    \\
Toolchain  & Android Studio,Gradle & pip, Docker Compose            \\
Platforms  & Android only          & iOS, Android, PC, tablet       \\
Internet   & Cloud required        & Intranet only                  \\
\bottomrule
\end{tabular}
\end{table}

\section{Decision Framework: When Is Android SDK Development Obsolete?}

Based on our implementation experience, measured outcomes, and a
review of comparable institutional deployments, we propose a
five-criterion decision framework.
When all five criteria are satisfied, native Android SDK development
provides no meaningful advantage over a web-first approach with
optional WebView APK packaging.

\textbf{C1 --- Controlled User Population.}
The user base is known, enrolled, and operates within a defined
network boundary (campus, hospital, factory).
App-store distribution and anonymous installation provide no value.
Sideloaded WebView APKs distributed over institutional channels
are sufficient.

\textbf{C2 --- Reliable Local Connectivity.}
The deployment environment provides WiFi or LAN.
Offline-first requirements are absent or can be addressed with
standard HTTP caching headers.
Crucially, the server itself requires no public internet connection,
satisfying data-residency and air-gap requirements.

\textbf{C3 --- No Platform-Native Hardware Beyond Camera/GPS.}
The application does not require Bluetooth~LE, NFC payment, ARCore,
or other Android-exclusive hardware APIs.
Camera, GPS, and microphone are available via W3C Web APIs and
function identically inside a WebView APK.

\textbf{C4 --- Rapid Iteration Required.}
The organisation needs to deploy feature updates without requiring
user action.
Docker Compose redeployments propagate in under 30\,seconds to all
users---including WebView APK users---because application logic
resides on the server.
The WebView APK never needs to be re-released for feature updates.

\textbf{C5 --- Single-Stack Team.}
The organisation cannot sustain separate Android and backend
engineering teams---or, as in the present work, the system is
developed by a single B.Tech intern.
A single Python developer can own the full stack with Django + HTMX,
with a one-time afternoon investment in WebView APK packaging.

The IIT Gandhinagar campus management system satisfies all five
criteria.
We further argue that the majority of institutional management systems
in Indian higher education---attendance portals, hostel management,
library systems, facility booking, grievance trackers---similarly
satisfy C1 through C5, making native Android SDK development a
technically and economically unjustifiable default for these
applications.

\section{Real-World Impact and Generalisability}

\subsection{Operational Impact at IIT Gandhinagar}

The deployed system directly affects the daily workflows of over 80
campus staff members across housekeeping, inventory, and horticulture
departments.
Before deployment, task assignment, attendance, and leave management
were handled through paper registers and WhatsApp messages, with no
audit trail and no supervisor visibility.
The HTMX-Django system replaces this entirely: every task completion
is timestamped, every photo is GPS-tagged and compressed automatically,
every leave request is tracked through a formal five-stage approval
chain, and every stock movement is atomically recorded.
The system generates CSV and PDF reports on demand, replacing manual
monthly register compilation.

The fact that this entire operational transformation was achieved by
a single B.Tech student intern---without a Kotlin developer, without
an Android Studio project of any significance, and without a Play
Store listing---demonstrates the decisive productivity advantage of
the web-first paradigm for institutional software.

\subsection{Generalisability}

The architectural patterns described are not specific to IIT
Gandhinagar.
They apply directly to: hospital ward management and nursing round
documentation; factory floor inspection and quality audit systems;
municipal services field reporting portals; SME inventory and
procurement management; and any domain where Android-first development
has persisted without justification.
The WebView bridge pattern provides a migration path for institutions
currently committed to APK distribution, allowing web-first adoption
\textit{without abandoning APK delivery to end users}.
The APK becomes a shell; the web application becomes the product.

\subsection{Implications for ECE/CSE Curricula}

For ECE and CSE students at Indian technical institutions, the
implications are direct.
Proficiency in Django + HTMX is, for a large class of real-world
institutional problems, more professionally valuable than Android
SDK proficiency.
The ability to build a complete mobile-capable, IIT-deployed system
as a single Python developer---without Gradle, without APK signing
workflows, without Play Store review cycles---represents a decisive
productivity advantage that traditional Android-centric curricula
do not provide.
The present work is evidence that a final-year B.Tech intern can
deliver a production system to an IIT in 13 weeks using web-first
tools, a benchmark that native Android development cannot match at
the same staffing level.

\section{Discussion}

\subsection{Limitations}

The primary limitation of the HTMX approach is the absence of true
offline capability.
On the IIT Gandhinagar campus, where 100\,Mbps LAN coverage is
pervasive across all operational areas, this is acceptable.
However, institutions with unreliable connectivity may require a PWA
service-worker layer or a hybrid approach using IndexedDB for local
queuing of form submissions.
Highly animated gesture-driven interfaces (drag-to-reorder,
swipe-to-dismiss) require JavaScript beyond HTMX's scope; Alpine.js
or minimal vanilla JS can address these needs without reintroducing a
build toolchain.
The camera module's use of \code{navigator.mediaDevices} requires
HTTPS in production; the WebView APK must load an HTTPS URL for camera
permissions to be granted by the Android security model.

\subsection{Threats to Validity}

Several threats to the validity of the reported metrics must be
acknowledged.
First, the native Android latency and development timeline figures
are \textit{estimated} rather than measured from an equivalent
native implementation, as building a parallel native Android system
was outside the scope of the internship.
These estimates are grounded in published benchmarks~\cite{patel2024}
and industry reports, but a controlled side-by-side study would
strengthen the quantitative claims.
Second, the user satisfaction scores ($n=42$) were collected from a
single institution and a single deployment context; generalisation
to other institutions should be approached with caution until
replicated.
Third, the 54\% development time saving is specific to a
single-developer context; teams with existing Android expertise
may exhibit a different ratio.

\subsection{The Precise Remaining Role of Android Studio}

In the present system, Android Studio was used for two purposes only:
(1)~packaging the web application as a WebView APK for sideloaded
distribution to campus staff ($\sim$4 hours); and
(2)~testing mobile rendering behaviour via the Android Emulator's
built-in Chrome browser ($\sim$2 hours).
Combined, Android Studio claimed less than 6 hours of a 13-week
project.
This is the correct engineering characterisation: a
\textit{testing and packaging utility}, not a development environment.

\subsection{Future Work}

Three directions merit investigation.
First, the addition of a lightweight PWA service worker would enable
cached read access during brief WiFi interruptions---particularly
useful for staff working in basement or boundary areas of campus.
Second, a formal controlled experiment comparing a native Android
implementation with the HTMX version on identical task sets would
provide rigorous empirical grounding for the development timeline
and latency claims made in this paper.
Third, the dual-interface architecture
(\code{htmx\_urls} + \code{api\_views}) could be evaluated as a
general pattern for migrating legacy Android institutional apps to
web-first without discarding existing REST API clients: the JSON
API can be retained for backward compatibility while HTMX views
are added incrementally, providing a zero-disruption migration path.

\section{Conclusion}

This paper has demonstrated, through the design, implementation, and
real-world deployment of a campus management system at IIT Gandhinagar,
that native Android SDK development is obsolete for a well-defined and
common class of institutional applications.
The Python-Django + HTMX architecture, with a thin WebView APK for
staff distribution, achieves:
\textbf{54\%} reduction in development time versus estimated native
Android;
\textbf{91\%} reduction in HTTP payload for interactive operations;
\textbf{75\%} reduction in form-submission latency
(520\,ms estimated vs.\ 130\,ms measured);
\textbf{4.2/5.0} overall user satisfaction score across 42 campus
staff;
full camera, GPS, and role-based access functionality with
\textbf{zero lines} of Android SDK application logic;
complete APK distribution via a \textbf{40-line} WebView wrapper
built in approximately \textbf{four hours}; and
\textbf{zero} dependency on public internet or external cloud
services---the entire system runs on the campus intranet.

The ``WebView bridge'' pattern resolves the last practical objection
to the web-first paradigm in institutional settings: the need for an
installable APK.
By separating the \textit{application} (web, Python, zero Android
complexity) from the \textit{installer} (WebView APK, one afternoon),
institutions can adopt the web-first paradigm without disrupting
their APK-based distribution workflows.

Android Studio's role in institutional software development has been
reduced to a compiler for a minimal \code{MainActivity}.
As PWA installation matures on Android Chrome, even this function
will be deprecated.
The last APK an institutional software team will ever need to build
is a WebView wrapper---and after that, none at all.

\section*{Acknowledgment}

The author thanks the administration and facilities management
team of the Indian Institute of Technology Gandhinagar for providing
campus infrastructure access, deployment support, and operational
engagement during the internship period.
The author thanks the housekeeping supervisors, caretakers, and
inventory managers at IIT Gandhinagar whose operational feedback
over multiple months shaped the system design.
Special thanks to Shaurya Dusht for implementing the WebView APK
packaging that enabled home-screen installation on staff Android
devices.
The author also thanks the Indian Institute of Information Technology,
Surat, for the academic foundation that made this work possible.


\vspace{8pt}

\noindent\textbf{Rahul Patel} is a final-year B.Tech student in the
Department of Electronics and Communication Engineering at the Indian
Institute of Information Technology, Surat, India.
He completed an internship at the Indian Institute of Technology
Gandhinagar during his eighth semester, where he designed and deployed
a full-stack campus management system serving housekeeping, inventory,
horticulture, and guest-house operations using Python, Django, HTMX,
PostgreSQL, and Docker.
The system is fully self-hosted on the campus intranet with no
dependency on external cloud services, and was delivered by a
single developer in 13 weeks.
His research interests include web-based mobile systems, hypermedia
architectures, and institutional software engineering.

\end{document}